\documentclass[sigconf]{acmart}
\usepackage{amsmath}
\usepackage{algorithm}
\usepackage{algorithmic}
\setcopyright{none}
\renewcommand\footnotetextcopyrightpermission[1]{} 
\AtBeginDocument{%
  \providecommand\BibTeX{{%
    \normalfont B\kern-0.5em{\scshape i\kern-0.25em b}\kern-0.8em\TeX}}}

\begin{document}

\title{Distributed Node Covering Optimization for Large Scale Networks and Its Application on Social Advertising}

\author{Qiang Liu}
\email{Q.L.Liu@hotmail.com}
\orcid{0000-0003-3171-1406}
\affiliation{%
  \institution{Tencent Corporation}
  \streetaddress{15 Ke Yuan Road}
  \city{Shenzhen}
  \country{China}
  \postcode{518057}
}


\begin{abstract}
    Combinatorial optimizations are usually complex and inefficient, which limits their applications in large-scale networks with billions of links. We introduce a distributed computational method for solving a node-covering problem at the scale of factual scenarios. We first construct a genetic algorithm and then design a two-step strategy to initialize the candidate solutions. All the computational operations are designed and developed in a distributed form on \textit{Apache Spark} enabling fast calculation for practical graphs. We apply our method to social advertising of recalling back churn users in online mobile games, which was previously only treated as a traditional item recommending or ranking problem.
\end{abstract}

\maketitle

\section{Introduction}
Many large systems can be modelled as graphs, such as the Internet \cite{faloutsos1999power}, brain \cite{20161}, and human \cite{10.2307/3629752} (or animal \cite{10.1093/comnet/cnab001}) social connectives. A graph consists of nodes and links. For a social structure, a node denotes a person while a link denotes the existence of a relationship between two nodes. The social system represented by a graph is usually called a social network \footnote{In this paper, we mix using the terms \textit{graph} and \textit{network}.}. The social network in an online platform is straightforward: a node is a user while a link means that two nodes are connected in the platform, i.e., two users can present messages to each other. 

People use online social network services (SNS) to communicate. In the meantime, SNS becomes also a cheap (perhaps even free) advertising (AD) medium by promoting users spreading ADs. In this brief paper, we discuss a graph optimization problem from an actual application aiming to optimize the effectiveness of reaching churn users through SNS ADs for online mobile games.

\section{Reaching Churn Users Through Active Ones}
\label{sec:churnCon}
Online games lose many users during the operation period. To maintain an instant number of daily active users, game companies usually advertise their game product through traditional media by directly presenting the game to the churn users. However, the expense for direct advertising is higher compared to reaching churn users through free-of-charge SNS channels among players of the game. 

\begin{figure}[htbp]
  \centering
  \includegraphics[width=\linewidth]{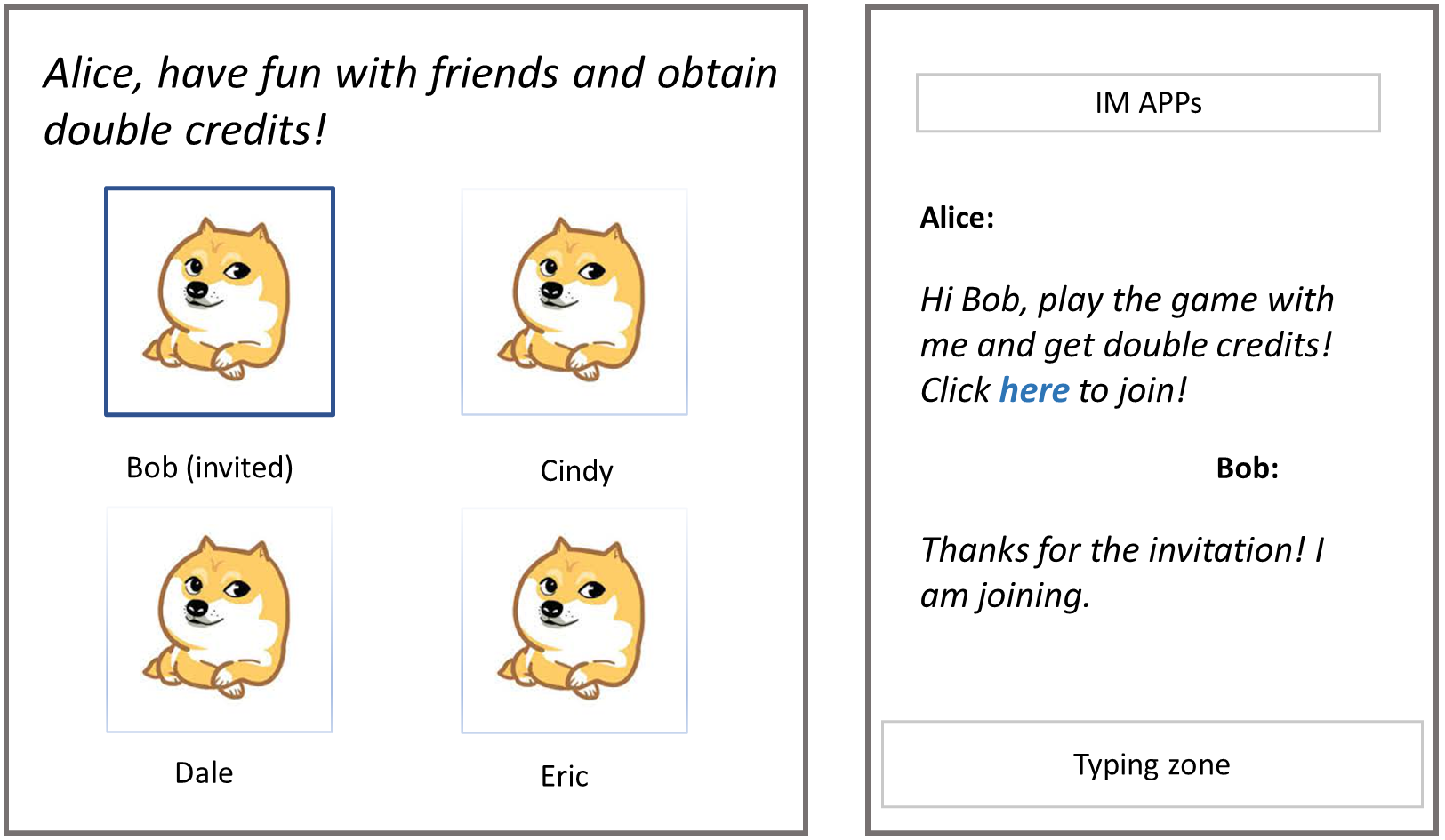}
  \caption{Example user interfaces of SNS ADs to reach churn users: the left interface appears in the mobile game APPs and the right one is the IM APP.}
  \label{fig:recall}
\end{figure}

A typical SNS AD to recall churn users of a mobile game is as follows. Users register the game through their SNS accounts, e.g. Instant Message (IM) APPs, and grant access to their IM contact list which contains the users' friends who are also the users of the game. In the game APP, an advertising module presents churn friends to active users and promotes them to recall their churn friends back. In most cases, active users can easily click the icons of their friends and the game APP automatically sends a message through the IM API to the user who has been clicked. As shown in Fig.~\ref{fig:recall}, active user Alice viewed the AD page presenting his churn friends of the game. Alice then clicked her friend Bob's icon and the system sent a message through Alice's IM to Bob. Bob received the message and can join the game as an active user through the link in the message.

\begin{figure}[htbp]
  \centering
  \includegraphics[width=\linewidth]{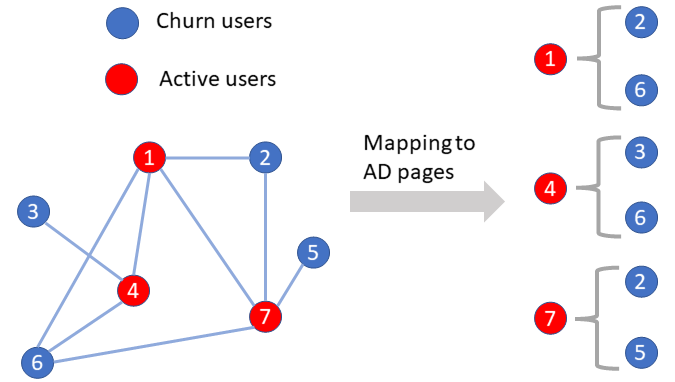}
  \caption{The mapping between the social network and the friend recommend list.}
  \label{fig:mapping}
\end{figure}

For each active user, presenting the churn friends is treated as a traditional item recommending or ranking optimization problem \cite{zhang2019deep}. Different strategies are applied for ranking, such as ranking more intimate friends higher, ranking users who are most likely to return higher, and ranking by machine learning models trained by user behaviors from past AD logs. However, those ranking methods may obtain an optimal result for each active user but cannot reach a global optimum of the total number of churn users being reached: churn users can be reached redundantly due to the overlapping of active friends in the social network. Fig.~\ref{fig:mapping} shows the mapping between the IM social network where neighbors can message each other and the churn friend recommending lists on the AD pages for active users. Users 1, 4, and 7 can reach all $4$ churn users and two active users are planned to receive ADs. There are only $3$ churn users are presented if selecting users $\{1, 7\}$ or $\{1, 4\}$ but $4$ churn users if selecting $\{4, 7\}$. The social network leverages the access of churns users through reachable active users if we can globally optimize the node covering, which is never considered before in the SNS AD. Furthermore, the problem to be solved in this paper is also different from other methods such as identifying the important/influential nodes or the key opinion leaders in networks \cite{bian2019identifying}, since those methods still achieve only local optima in the SNS AD problem.

To achieve optimal access to churn users, we need to find a subset of the active users whose churn neighbor node set is the largest, i.e. a maximum node covering. Let $\mathcal{V}_a$ and $\mathcal{V}_c$ denote the node sets of active users and churn users, respectively. Let $\mathcal{E}_{ac}$ denote the link set and each link connects an active user and a churn user. Graph $G=(\mathcal{V}_a, \mathcal{V}_c, \mathcal{E}_{ac})$ is thus a bipartite graph: there is no link within $\mathcal{V}_a$ and $\mathcal{V}_c$, i.e. $\mathcal{E}_{ac}\subseteq\{(v_a, v_c)| v_a\in\mathcal{V}_a, v_c\in\mathcal{V}_c\}$. The goal is to find the optimal solution
\begin{equation}
V_a^*\triangleq\operatorname*{argmax}_{V_a \subseteq \mathcal{V}_a, |V_a|=N}|\{v_c|v_c\in V_c, v_a\in V_a, (v_a, v_c)\in \mathcal{E}_{ac}\}|
\label{eq:eq1}
\end{equation}
where $N$ is the number of active users receiving the SNS AD related to the budget. The particular combinatorial problem (\ref{eq:eq1}) is similar to many other similar problems of vertex covering or matching in graphs \cite{APOLLONIO201437,CHEN2003833} with sometimes NP-Hard complexity \cite{10.2307/3597377}. Classical approximation algorithms are not feasible to be applied in real social networks. Although heuristic solutions are reported to produce better results with high efficiency to some node-covering problems compared to approximation algorithms \cite{4424701}, the studied graphs are still tiny pieces compared to those in practice. In one of our deployments, there are \textasciitilde$170$ millions of nodes and \textasciitilde$1.7$ billions of links. A distributed algorithm is needed to solve Eq.~(\ref{eq:eq1}) at such a large scale.

\section{A genetic algorithm for node covering}
A genetic algorithm (GA) \cite{4424701} mimics the evolution of biological species and explores the solution space toward a pre-defined optimal goal. In a normal GA, there are four key terms: \textit{group}, \textit{individual}, \textit{gene}, and \textit{fitness}. A group is a set of individuals and an individual is a set of genes. The fitness of an individual is a metric that evaluates the quality of individuals, i.e., solutions to the defined problem. In addition, there are three operators that mimic the biological evolutionary dynamics: \textit{selection}, \textit{crossover}, and \textit{mutation}. \textit{Selection} applies to each candidate individual in the group by randomly keeping or discarding the individual at each iteration. \textit{Crossover} applies to two individuals within the group by randomly exchanging part of their genes. \textit{Mutation} randomly replaces a part of the genes in an individual by randomly selected genes from the whole gene set. Selection and crossover aim to produce better individuals while mutation is to keep the diversity of the individual and explore the solution space.

\begin{figure}[htbp]
  \centering
  \includegraphics[width=\linewidth]{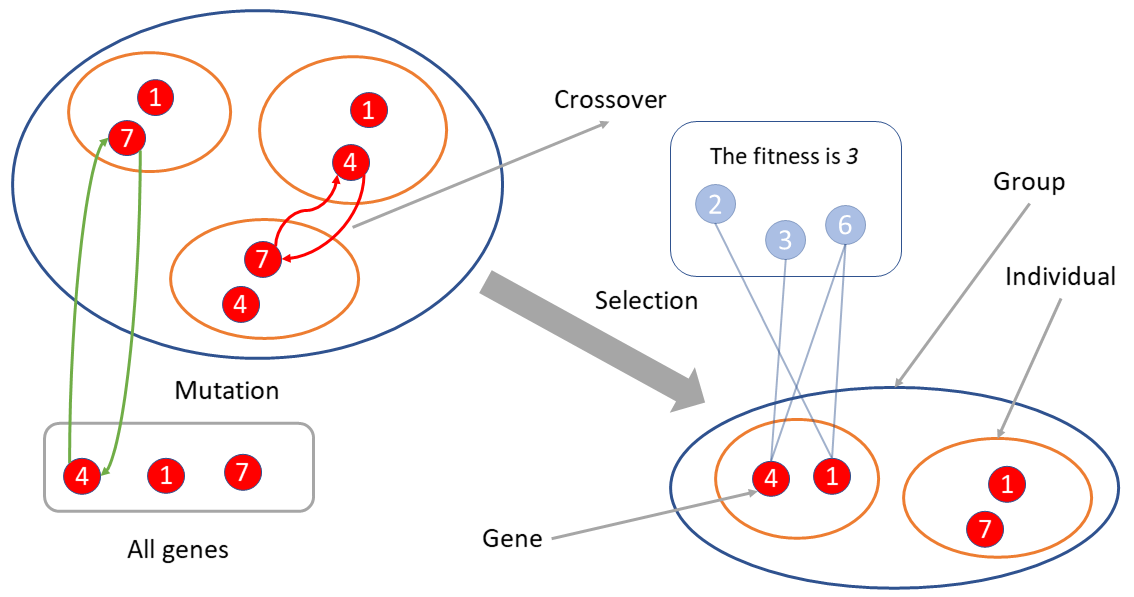}
  \caption{The GA design for our node covering problem.}
  \label{fig:definition}
\end{figure}

For our SNS ADs, we need to map the node-covering problem to GA terminologies. As shown in Fig.~\ref{fig:definition}, we define nodes $v_a\in\mathcal{V}_a$ denoting the active users as genes and thus, an individual of GA is a subset $V_a \subseteq \mathcal{V}_a$ of active user nodes. The group is just a candidate individual set $S\subseteq\{V_a|V_a\subseteq \mathcal{V}_a\}$. Our goal is to find the optimal individual $V_a^*\subseteq \mathcal{V}_a$ through the evolving group $S$ as in Eq.~(\ref{eq:eq1}), which is the target active users of the SNS AD. The quality of an individual is measured by fitness, which is a mapping $\mathcal{F}: \{V_a |V_a\subseteq \mathcal{V}_a\} \rightarrow \mathcal{R}$. As a result of Eq.~(\ref{eq:eq1}), the fitness can be defined as 
\begin{equation}
    f(V_a) \triangleq |\{v_c|(v_a,v_c)\in \mathcal{E}_{ac}, v_a\in V_a\}|.
    \label{eq:fitness1}
\end{equation}
Definition~(\ref{eq:fitness1}) may not be optimal in real scenarios, because the churn user nodes in $V_c$ are not equivalent due to their diverse user features. We will discuss another definition fitting in practice later in Sec.~\ref{sec:nodemeasure}. As shown in Fig.~\ref{fig:definition}, the individual $\{1, 4\}$ has $3$ neighbors $\{2,3,6\}$ and thus fitness $f(\{1, 4\})=3$.

After all the mapping from our problem to the GA framework, the GA procedure is straightforward by applying the genetic operators to the group $S$ as described in Algo.~\ref{algo}. GA searches the optimal solution $V_a^*$ by producing possible candidate solutions through crossovers and exploring the solution space by mutations. At each iteration, individuals with higher fitness are more likely to be preserved by selection. In the mutation step of Algo.~\ref{algo}, the gene removal and adding rates are set to be $M$ and $\frac{MN}{|\mathcal{V}_a|-N}$, respectively, to maintain a constant number of genes in each individual.
 
\begin{algorithm} 
	\caption{GA to solve $V_a^*=\operatorname*{argmax}_{V_a}f(V_a)$} 
	\label{algo} 
	\begin{algorithmic}
		\REQUIRE Graph $G=(\mathcal{V}_a,\mathcal{V}_c,\mathcal{E}_{ac})$; Crossover probability of individuals $C$; Mutation probability of genes $M$; Number of genes in each individual $N$; Number of individuals of the group $T$; Number of iterations $I$.
		\ENSURE $V_a^*$ 
		\STATE \textbf{Initialize:} Construct the initial group containing $T$ individuals $S\subseteq\{V_a|V_a\subseteq\mathcal{V}_a, |V_a|=N\}$ and a next generation group $S'=\emptyset$.
		\FOR{$i=1$; $i\leq I$; $i++$}
		\FOR{$j=1$; $j\leq T$; $j++$}
		\STATE \textbf{Calculate fitness:} For each $V_a\in S$, calculate individual fitness $f(V_a)$ by Eq.~(\ref{eq:fitness1}).
		\STATE \textbf{Selection: } Select two individuals $V'_a$ and $V''_a$ from $S$ with probability $\Pr(V_a)=\frac{f(V_a)}{\sum_{V'_a\in S}f(V'_a)}$; 
		\STATE \textbf{Crossover: } Exchange genes of $V'_a$ and $V''_a$ with probability $C$ and Boolean indicators of each gene are exchanged with probability $p$, where $p\sim U(0, 1)$;
		\STATE \textbf{Mutation: } Remove each gene $v'_a\in V'_a$ from $V'_a$ with probability $M$ and for each gene $v_a\in V'_a\backslash\mathcal{V}_a$, add $v_a$ into $V'_a$ with probability $\frac{MN}{(|\mathcal{V}_a|-MN)}$.
		\STATE Let $S' \leftarrow S'\cup \{V'_a\}$. 
		\ENDFOR
		\STATE Let $S\leftarrow S'$ and $S'\leftarrow\emptyset$. 
		\ENDFOR
	\STATE Let $V_a^*=\operatorname*{argmax}_{V_a\in S}f(V_a)$.
	\end{algorithmic} 
\end{algorithm}

\section{A Two-step Strategy for Constructing the Initial Group $S$}
\label{sec:twostep}
To ensure Algo.~\ref{algo} has an efficient start, we design a two-step strategy to construct initial individuals. An ideal initial group $S$ should have two features: 1) Individuals in $S$ contain diverse genes to avoid GA ending up into local optima of the solution space; 2) The expected fitness of each individual is as high as possible for an efficient convergence to optima. We apply the following two-step procedure to construct the initial group:
\begin{enumerate}
\item  For each churn user node $v_c\in \mathcal{V}_c$, one and only one of it active neighbor $v_a\in\mathcal{V}_a$ is preserved in set $V_i$;
\item  Select genes only from $V_i$ to construct the initial individuals $V_a\in S$.
\end{enumerate}
Step 1 ensures that the candidate genes in $V_i$ cover all the churn user nodes in $\mathcal{V}_c$. In both steps 1 and 2, greedy and random selecting strategies are applied to preserve genes. For example in step 1, we can preserve the one neighbor $v_a$ for each $v_c\in V_c$ with the largest fitness $\max_{v_a\in\{v_a|(v_a, v_c)\in\mathcal{E}_{ac}\}}f(\{v_a\})$ to obtain excellent candidate genes. However, this greedy preserving strategy may result into local optima since a combination of nodes $v_a$ with large fitness may redundantly connect overlapping $v_c\in\mathcal{V}_c$. By selecting $v_a\in\mathcal{V}_a$ with smallest fitness $f(\{v_a\})$, the redundancy of common neighbors is reduced and more candidate genes can be obtained in $V_i$. Similar conclusions apply to step 2. The combination of three strategies of selecting $v_a\in\mathcal{V}_a$ in step 1 and 2:
\begin{itemize}
    \item selecting with probability proportional to $f(\{v_a\})$; 
    \item selecting with probability inversely proportional to $f(\{v_a\})$; 
    \item selecting randomly irrelevant to $f(\{v_a\})$, 
\end{itemize}
enables of constructing initial individuals with both high diversity of genes and relatively large fitness, which boosts the efficiency of our GA. 

\section{A Distributed GA Realization}
\label{sec:distribute}
Algorithm~\ref{algo} cannot directly find the optimal active user sets $V_a^*$ for our applications because practical networks are at the scale of billions. To facilitate the optimization of our SNS ADS, we realize Algo.~\ref{algo} on \textit{Apache Spark} \citep{10.1145/2934664} distributively. For each task on Spark, a large number of executors collaboratively schedule the calculation and each executor is only responsible for the computation task of a part of the whole dataset. The basic data object to be operated on Spark is the Resilient Distributed Dataset (RDD), which represents the actual data distributively stored. The structure of RDDs is similar to a table: each row represents a data sample and each column represents a field of the samples. 

For our realization, the RDD structure is described in Table~\ref{tab:RDD}. There are in total $2T+2$ columns and $|\mathcal{V}_a|$ rows in the table. Each row represents a gene of active user $v_a\in\mathcal{V}_a$. The last $2T$ columns represent the $T$ individuals in the group $S$ (columns $1$ to $T$) and $S'$ (columns $T+1$ to $2N$); Each element in the first column is the active user node $v_a\in\mathcal{V}_a$ while the second column stores the neighbors $v_c\in\mathcal{V}_c$ of $v_a$. The elements of column \textit{Individual $K$} indicate whether the nodes $v_a$ of the first column are contained in individual $k$ or not by a Boolean value. 

In Spark, the RDD rows representing genes are stored in different executors and all the operators (selection, crossover, and mutation) can be applied independently, and thus distributively, on each row. \textit{Selection} copies individual $1\leq k\leq T$ columns into individual $T+k$ columns, \textit{crossover} exchanges elements between two individual columns, and \textit{mutation} flips Boolean elements of individual columns. 

Calculation of fitness of each individual is realized first by Spark operator \textit{filter} which preserves all the genes with Boolean indicators being $1$ in the corresponding column. And then, count the number of different nodes $v_c$ in the whole RDD, which is realized by applying Spark operator \textit{flatMap} and \textit{reducedByKey} on the second column. \textit{FlatMap} splits each Array in the second column into elements and flattens the elements $v_c$ into single rows. The operator \textit{reduceByKey} merges duplicated rows of $v_c$ into one row. Thus, we can just obtain the fitness of the calculated individual by counting the number of rows in the resultant RDD. 

For the two-step strategy in Sec.~\ref{sec:twostep}, the input RDD is an edge list with each row being $(v_a, v_c)$. The first step is realized by applying Spark operator \textit{groupBy} on each element $v_c$ in the second column and the second step is by assigning a Boolean indicator for each selected gene. 

\begin{table*}
  \caption{The example RDD structure for Algo.~\ref{algo} corresponds to the graph shown in Fig.~\ref{fig:mapping}.}
  \label{tab:RDD}
  \begin{tabular}{cccccc}
    \toprule
    Node $v_a\in \mathcal{V}_a$ (big int) & Neighbors $v_c$ of $v_a$ (Array) & Individual $1$ & Individual $2$ & $\cdots$ & Individual $2N$ \\
    \midrule
    1 & [2, 6] & 1 & 1 & $\cdots$ & 0\\
    4 & [3, 6] & 0 & 1 & $\cdots$ & 1\\
    7 & [2, 5] & 1 & 0 & $\cdots$ & 1\\
    \bottomrule
  \end{tabular}
\end{table*}

With all the above-mentioned data structure and the operating procedure designs, we apply our distributed GA on a network with $\sim 170$ million nodes and $\sim 1.7$ billion links. The limiting number of genes in each individual is $9$ million. Table~\ref{tab:performance} shows the GA performance compared to the degree (the number of churn-friend nodes) greedy and random selecting methods with the coverage increased by $10.15\%$ and $15.90\%$, respectively. 
\begin{table}
  \caption{The number of covering nodes $v_c\in\mathcal{V}_c$ by $V_a^*$.}
  \label{tab:performance}
  \begin{tabular}{ccc}
    \toprule
    Method & \# Churn nodes covered & GA increased by  \\
    \midrule
    GA & $48805199$ & $-$ \\
    Degree greedy & $44306059$ & $10.15\%$ \\
    Random & $42110428$ & $15.90\%$ \\
    \bottomrule
  \end{tabular}
\end{table}
\section{Design of the Fitness in Real ADs}
\label{sec:nodemeasure}
In real scenarios, nodes $v_c\in\mathcal{V}_c$ are not equivalent. Churn users re-login to the game with diverse probabilities after being reached by the SNS AD. Some churn users are lost forever and will never come back. Thus, each node $v_c\in\mathcal{V}_c$ should be measured by its return probabilities. Moreover, a node $v_c$ is reached with variety of probabilities through active nodes in $\mathcal{V}_a$ and some active users may never click the SNS AD. The fitness defined in Eq.(\ref{eq:fitness1}) lacks measuring those user behavioral diversities.

We define a measure $m(\bullet)$ for each link $(v_a, v_c) \in \mathcal{E}_{ac}$ which incorporates the user behaviors by
\begin{equation}
    \label{eq:measure}
    m(v_a, v_c) \triangleq \Pr(v_a\ {\rm clicks}\ v_c, v_c\ {\rm returns}).
\end{equation}
Eq.~(\ref{eq:measure}) is not directly calculable but can be estimated with log data from previous AD sessions. With historical user behaviors and features, we train an XGBoost model \cite{10.1145/2939672.2939785} to predict the measure (\ref{eq:measure}) of each node pair $(v_a,v_c)\in \mathcal{E}_{ac}$ for new AD sessions.

The improved fitness of individuals $V_a\subset\mathcal{V}_a$ is defined as follows
\begin{equation}
    \label{eq:fitness2}
    f(V_a) = \sum_{v_c\in S_c(V_a)}\frac{\sum_{v_a\in S_{V_a}(v_c)}m(v_a, v_c)}{|S_{V_a}(v_c)|}
\end{equation}
where $S_c(V_a) = \{v_c|(v_a, v_c)\in \mathcal{E}_{ac},~v_a\in V_a\}$ is the set of churn nodes covered by $V_a$ and $S_{V_a}(v_c) = \{v_a|(v_c, v_a)\in \mathcal{E}_{ac},~v_a\in V_a\}$ is the set of active nodes in $V_a$ connected to the specified $v_c$. Compared to Eq.~(\ref{eq:fitness1}), Eq.~(\ref{eq:fitness2}) first calculate the mean measure of each churn node $m(v_a, v_c)$ over $S_{V_a}(v_c)$ and then do the similar fitness calculation as in (\ref{eq:fitness1}) over $V_c\subseteq \mathcal{V}_c$ that covered by $V_a$. If we set $m(v_a,v_c)=1$, then the right terms after the summations are $1$ and Eq.~(\ref{eq:fitness1}) and Eq.~(\ref{eq:fitness2}) become equivalent. 

The computation of fitness (\ref{eq:fitness2}) is realized by simply passing a sum function and a counting into the \textit{reduceByKey} operator when merging $v_c$, followed by a mean calculation mapped to each row of the flatten RDD. Finally, sum over the fitness of each $v_c$ in the resultant RDD similar to that in Sec.\ref{sec:distribute}. 

Some miscellaneous improvements are also applied in practice to Algo.~\ref{algo}. At each iteration, we preserve the individual with the highest fitness to avoid losing the best solution during GA running. When calculating the probability of selecting individual $\Pr(V_a)$ in the \textit{selection} step, we subtract a base number from each $f(V_a)$ to avoid the small deviating fitness among individuals.  

Assisted by the above-mentioned improvements, GA has achieved considerable large gains in conversion rate (\# inactive-to-active over \# exposes) compared to non-optimized scenarios. As one typical example, the conversion rate has increased by $10.13\%$ for the out-of-game channel and $65.70\%$ for the in-game channel in one of our games.

\section{Conclusion}
In this brief report, we introduce a distributed computing scheme to solve a combinatorial optimization problem on large-scale networks. The problem is formulated from real scenarios in mobile-game ADs, where churn users are advertised through active users on social networks. This optimization problem was generally not feasible at billion-scale graphs in practice and the scene was only treated as a traditional item recommending/ranking problem before. Our distributed GA achieves a considerable large performance gain with high efficiency in our SNS AD scenes.
\begin{acks}
We thank Dr. Shenggong Ji for helpful discussions during the realization of the project.
\end{acks}

\bibliographystyle{unsrt}
\bibliography{sample-base}

\appendix

\end{document}